\documentclass{article}

\usepackage{arxiv}

\usepackage[utf8]{inputenc} 
\usepackage[T1]{fontenc}    
\usepackage{hyperref}       
\usepackage{url}            
\usepackage{booktabs}       
\usepackage{amsfonts}       
\usepackage{nicefrac}       
\usepackage{microtype}      
\usepackage{lipsum}
\usepackage{graphicx}
\usepackage{amsmath}
\graphicspath{ {./images/} }

\title{Congenital Heart Disease Classification Using Phonocardiograms: A Scalable Screening Tool for Diverse Environments}

\author{
 Abdul Jabbar \\
  Department of Electrical and Computer System Engineering\\
  Monash University\\
  Clayton, VIC 3800 Australia \\
  \texttt{abdul.jabbar@monash.edu} \\
   \And
 Ethan Grooby \\
  Department of Biomedical Engineering\\
  McGill University\\
  Quebec, Canada  \\
   \And
 Jack Crozier \\
  Department of Electrical and Computer System Engineering\\
  Monash University\\
  Clayton, VIC 3800 Australia  \\
   \And
 Alexander Gallon \\
  Department of Electrical and Computer System Engineering\\
  Monash University\\
  Clayton, VIC 3800 Australia  \\
   \And
 Vivian Pham \\
  Department of Electrical and Computer System Engineering\\
  Monash University\\
  Clayton, VIC 3800 Australia  \\
   \And
 Khawza I Ahmad \\
  Department of Electrical and Electronic Engineering\\
  United International University\\
  Dhaka, Bangladesh  \\
   \And
 Raqibul Mostafa \\
  Department of Electrical and Electronic Engineering\\
  United International University\\
  Dhaka, Bangladesh  \\
   \And
 Md Hassanuzzaman \\
  School of Engineering\\
  Duke University\\
  USA  \\
   \And
 Ahsan H. Khandoker \\
  epartment of Biomedical Engineering\\
  Khalifa University\\
  Abu Dhabi 127788, UAE\\
   \And
 Faezeh Marzbanrad \\
  Department of Electrical and Computer System Engineering\\
  Monash University\\
  Clayton, VIC 3800 Australia \\
  \texttt{faezeh.marzbanrad@monash.edu}\\
}
\begin{document}
\maketitle
\begin{abstract}
Congenital heart disease (CHD) is a critical condition that demands early detection, particularly in infancy and childhood. This study presents a deep learning model designed to detect CHD using phonocardiogram (PCG) signals, with a focus on its application in global health. We evaluated our model on several datasets, including the primary dataset from Bangladesh, achieving a high accuracy of 94.1\%, sensitivity of 92.7\%, specificity of 96.3\%. The model also demonstrated robust performance on the public PhysioNet Challenge 2022 and 2016 datasets, underscoring its generalizability to diverse populations and data sources. We assessed the performance of the algorithm for single and multiple auscultation sites on the chest, demonstrating that the model maintains over 85\% accuracy even when using a single location. Furthermore, our algorithm was able to achieve an accuracy of 80\% on low-quality recordings, which cardiologists deemed non-diagnostic. This research suggests that an AI-driven digital stethoscope could serve as a cost-effective screening tool for CHD in resource-limited settings, enhancing clinical decision support and ultimately improving patient outcomes.
\end{abstract}

\keywords {Echocardiography \and Phonocardiography (PCG)\and Deep Learning, Congenital heart disease (CHD) \and Low- and Middle-Income Countries (LMIC)}

\section{Introduction}
\label{sec:introduction}
Congenital heart disease (CHD), also known as congenital heart defects, occurs in approximately 1.2\% of newborns globally, with one in four cases being severe \cite{wu2020incidence}. At least 220,000 people die from this disease annually, with most deaths occurring within the first year of life \cite{zimmerman2020global}. In some extreme cases, CHD can damage the heart valves and limit blood circulation through the coronary arteries or other parts of the body, typically requiring surgery within the first year of life \cite{oster2013temporal}. While CHD mortality rates have dropped dramatically in high-income nations due to advances in diagnosis and treatment, it remains a major cause of morbidity and mortality in Low- and Middle-Income Countries (LMICs) \cite{zimmerman2020congenital}. These regions often lack the resources and advanced medical care necessary for CHD diagnosis and treatment, leading to higher costs and mortality rates associated with CHD \cite{rahman2019linking}. Early detection through emerging screening methods could potentially reduce mortality rates in these countries by about 58\%, preventing many cases of disability and death \cite{higashi2015burden}.

Echocardiography is currently the gold standard for detailed pre- and post-natal screening and is used to diagnose CHD \cite{lytzen2018live}. Although echocardiography is an established and reliable test, its high cost and the need for skilled specialists to operate and interpret the results make it difficult to implement in low-resource settings. Additionally, variations in professional interpretations \cite{mcleod2018echocardiography} among clinicians can delay the early detection of cardiac abnormalities when relying solely on expert opinion \cite{chorba2021deep}. Moreover, many patients in rural and regional areas often have to travel considerable distances to reach hospitals with the necessary resources and trained personnel \cite{hoodbhoy2021diagnostic}.

Phonocardiography (PCG), a non-invasive technique for cardiac auscultation that provides insights into the heart's mechanical function \cite{alkhodari2021convolutional}, is a potential alternative \cite{burns2022application}. Using an electronic stethoscope, PCG signals capture the activity of the mitral, tricuspid, aortic, and pulmonic valves, along with other cardiac structures, as blood flows through and around the heart \cite{aziz2020phonocardiogram}. Four key cardiac sounds are produced as blood passes through these valves: S1, S2, systole, and diastole.
Irregular heartbeats or heart murmurs during diastole or systole may indicate CHD \cite{sepehri2016intelligent}. Specifically, a diastolic heart murmur occurs after S2 and before S1, while a systolic murmur occurs between S1 and S2 \cite{wang2020automatic}.
With advancements in Artificial Intelligence (AI) and computerized algorithms, alongside the growing demand for accessible, continuous, and personalized healthcare, PCG could play a significant role in improving CHD diagnosis and management, particularly in LMICs \cite{bozkurt2018study}.


Automated assessment of PCG signals has gained increasing attention, culminating in the focus of the PhysioNet 2016 and 2022 Challenges on this topic. The objective of the PhysioNet Challenge 2016 was the automated classification of normal and abnormal adult PCG signals \cite{bib33}, while the objective of the PhysioNet Challenge 2022 was to automate the identification of murmurs and clinical outcomes from pediatric PCG signals \cite{bib34}. The top-performing team in the PhysioNet Challenge 2016 was Potes \textit{et al.}, who utilized AdaBoost and convolutional neural networks (CNN) for classifying heart sounds, extracting 124 features from the audio data, and achieving an accuracy of 86.02\%. Furthermore, the top-performing team in the PhysioNet Challenge 2022, Yujia Xu \textit{et al.}, achieved the highest weighted accuracy of 85.3\% in murmur detection using a hierarchical multi-scale CNN (HMS-Net). Their method involved extracting convolutional features from spectrograms at various scales and integrating them through a hierarchical structure \cite{xu2022hierarchical}. The second-place team, Hui Lu \textit{et al.}, developed a lightweight CNN and a Random Forest classifier for heart disease classification. Their approach utilized 128 mel spectrogram features from heart sounds, along with demographic information, achieving an weighted accuracy of 79.1\% in murmur detection \cite{lu2022lightweight}. The third-place team, McDonald \textit{et al.}, used a recurrent neural network combined with a hidden semi-Markov model (HSMM) for murmur detection, resulting in an weighted accuracy of 77.6\% \cite{mcdonald2022detection}. Al-Issa \textit{et al.} present a cardiac diagnostic system that combines CNN and long short-term memory (LSTM) components, achieving impressive accuracy rates of 98.5\% and 99.87\% for non-augmented and augmented datasets, respectively, while also demonstrating strong performance on the PhysioNet/Computing in Cardiology 2016 challenge dataset \cite{AlIssa2022}.

Despite these advances, there are significant research gaps in the practical implementation of detecting CHD using stethoscopes in LMICs. Most studies focus on murmur detection and CVD, which are not equivalent to CHDs. Aiming to detect CHDs using PCGs recorded in Bangladesh, Hassanuzzaman \textit{et al.}, proposed a transformer-based neural network, achieving  93\% accuracy and 97\% sensitivity \cite{bib45}. While proving the feasibility of detecting CHDs from heart sounds, did not address several practical challenges relevant to real-world deployment, particularly in LMICs. 
It is crucial to analyze how the algorithm performs with low-quality signals, that is a common issue in LMICs, especially in pediatric populations and mobile health contexts, where PCGs are often recorded by non-experts. Morover, the sensitivity of CHD detection to the precise site of sound collection on the chest also remains unexamined. It is unclear whether all four major auscultation sites are necessary for accurate CHD detection, or if a single location or subset of these would suffice. In a mobile health context, collecting PCGs from multiple precise locations could be challenging for non-experts, impacting the usability of the method. Furthermore, it has not been explored how well the algorithm, trained on a specific population and using a particular stethoscope, would generalize to a different population or device. The transferability of algorithms developed for other tasks, such as murmur classification, to CHD detection is also yet to be explored. Addressing these gaps is essential to ensure that CHD detection methods are robust, generalizable, and practical for use in resource-constrained settings.

These research gaps are addressed in this work. We propose a deep learning model to enable CHD detection using a digital stethoscope as a portable, low-cost method for mHealth applications in LMICs. The contributions of this paper are as follows:

\begin{itemize}
    \item Most databases and, consequently, most available algorithms are developed for murmur detection. However, murmurs do not necessarily indicate CHD. In this study, we introduce a robust deep learning algorithm specifically designed for detecting CHD in children using data collected in Bangladesh.
    \item We propose an algorithm that operates directly on raw PCG signals without requiring extensive preprocessing steps.
    \item We analyzed the transferability of the algorithm between different populations, stethoscopes, and tasks such as murmur detection.
    \item We identified the significance of each auscultation site on the chest for CHD detection.
    \item We evaluated the algorithm on PCG recordings flagged by cardiologists as too noisy or low quality, ensuring its robustness in real-world conditions.
\end{itemize}

\section{Method}
\begin{figure*}[ht]
    \centering
    \includegraphics[width=0.8\textwidth]{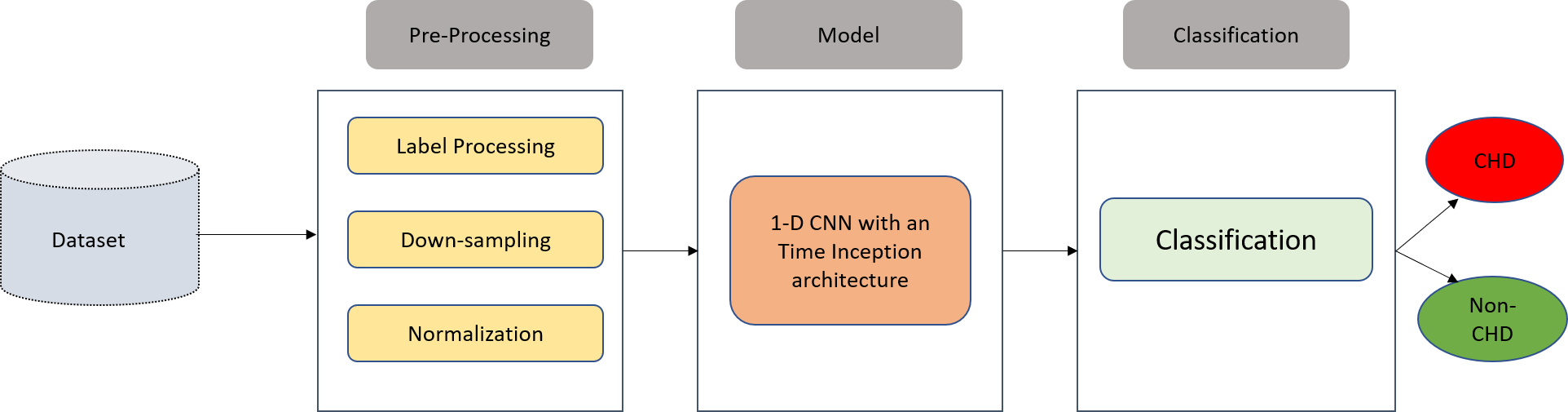}
    \caption{Block diagram of proposed model for CHD detection}
    \label{fig1:block1}
\end{figure*}
\subsection{Bangladesh Dataset}
\subsubsection{Dataset and Patient Enrolment}
The Eko DUO ECG + Digital Stethoscope was used to collect the PCG signals in Bangladesh Shishu (Children) Hospital \& Institute and National Heart Foundation Hospital \& Research Institute, which serves as the main dataset in this research. Data was collected from both healthy individuals (non-CHD) and those diagnosed with CHD in clinical settings. The project received human ethics approval from the regional medical ethics committees of the National Heart Foundation Hospital \& Research Institute (Ethics Approval Number: NHFH \& RI 4-14/7/AD/1132) and Bangladesh Shishu (Children) Hospital \& Institute (Ethics Approval Number: Admin/1714/BSHI/2021).

In this dataset, each patient had four or more consecutive PCG signals collected from the tricuspid valve (TV), pulmonary valve (PV), aortic valve (AV), and mitral valve (MV) auscultation sites. Each signal was collected for 15 seconds at a 4000 Hz sampling frequency. The dataset includes a total of 3,435 PCG signals from 751 individuals. Of these, 456 patients were diagnosed with CHD as confirmed by the cardiologists using chest X-rays, CT scans, 12-lead ECGs, echocardiograms, and clinical assessments such as auscultation of heart sounds. The remaining 295 patients were classified as not having any CHDs. As shown in the Tables \ref{tab:gender} and \ref{tab:population}, the dataset is skewed towards the male population, and most patients are in the age range of infancy to adolescence (5 months to 16 years).
Furthermore, samples from the National Heart Foundation Hospital \& Research Institute are predominantly from CHD patients, as these patients come to the hospital specifically for treatment. In contrast, samples from Bangladesh Shishu (Children) Hospital \& Institute include both CHD and non-CHD patients, as patients visit the hospital for CHD diagnosis as well as general care. The complete description of the dataset is shown in Tables \ref{tab:gender}, \ref{tab:population} \& \ref{tab:comparison}.
\subsubsection{Data Preparation and Labelling}
Each signal duration was standardized to 15 seconds. Longer recordings were trimmed, and shorter signals were zero-padded. Patient-wise labeling was extended to all recordings from multiple auscultation sites of the same individual, meaning that the same label (CHD or non-CHD) assigned to the patient was applied to all sounds recorded from different auscultation sites on their chest.
\begin{table*}[h]
    \centering
    \caption{gender distrubution}
    \begin{tabular}{|c|c|c|c|}
        \hline
        \textbf{} & \textbf{CHD, (\%)} & \textbf{Non-CHD, (\%)} & \textbf{Total, (\%)} \\
        \hline
        \textbf{Male} & 257 (56) & 190 (64) & 447 (60) \\
        \hline
        \textbf{Female} & 199 (44) & 105 (36) & 304 (40) \\
        \hline
    \end{tabular}
    \label{tab:gender}
\end{table*}

\begin{table*}[h]
\centering
\caption{Age distrubution }
\label{tab:population}
\begin{tabular}{|c|c|c|c|c|}
\hline
\textbf{Group} & \textbf{Age (Years)} & \textbf{CHD, N (\%)} & \textbf{Non-CHD, N (\%)} & \textbf{All, N (\%)} \\
\hline
\textbf{Infant}    & 1 month to 2 & 125 (27) & 137 (46) & 262 (35) \\
\hline
\textbf{Child}  & 2 to 12      & 306 (67) & 152 (52) & 458 (61) \\
\hline
\textbf{Adolescence} & 12 to 16     & 22 (5)   & 6 (2)    & 28 (4)   \\
\hline
\textbf{Adult}      & $>$ 16         & 3 (1)    & 0 (0)       & 3 (1)    \\
\hline
\textbf{Total} & & \textbf{456 (61)} & \textbf{295 (39)} & \textbf{751} \\
\hline
\end{tabular}
\end{table*}
\begin{table*}[h]
\centering
\caption{Comparison of Weight, Height, and BMI between CHD and Non-CHD Groups}
\label{tab:comparison}
\begin{tabular}{|c|c|c|c|c|c|c|}
\hline
\textbf{} & \multicolumn{2}{c|}{\textbf{Weight (kg)}} & \multicolumn{2}{c|}{\textbf{Height (cm)}} & \multicolumn{2}{c|}{\textbf{BMI}} \\
\hline
\textbf{} & \textbf{CHD} & \textbf{Non-CHD} & \textbf{CHD} & \textbf{Non-CHD} & \textbf{CHD} & \textbf{Non-CHD} \\
\hline
\textbf{Max}  & 62.5  & 80    & 242    & 177   & 27.34 & 29.6  \\
\hline
\textbf{Min}  & 3.4   & 2.3   & 58     & 54    & 4.54  & 4.7   \\
\hline
\textbf{Mean} & 15.98 & 15.18 & 103.93 & 98.4 & 14.02 & 14.61 \\
\hline
\textbf{STD}  & 8.61 & 10.21 & 24.59  & 26.95 & 2.38  & 3.42  \\
\hline
\end{tabular}
\end{table*}
\subsection{PhysioNet Challenge 2022 Dataset}
\subsubsection{Dataset and Patient Enrolment}
The CirCor DigiScope dataset from the George B. Moody PhysioNet 2022 Challenge was used for pre-training and evaluating transferability of the methods \cite{bib34}. As part of the 'Caravana do Coração' (Caravan of the Heart) program, pediatric patients from Northeast Brazil were enrolled in two mass screening campaigns that took place in July-August 2014 and June-July 2015. Approval for the data collection protocol was granted by the 5192-Complexo Hospitalar HUOC/PROCAPE institutional review board at the request of the Real Hospital Portugues de Beneficencia em Pernambuco. Consent forms were signed by all individuals aged 21 years or under; for those under 18, consent forms were provided by legal guardians.

There were 1,568 patients in the trial, and 942 of them (or around 60\%) had data that was made publicly available. One or more PCG signals were collected sequentially from the AV, MV, TV, and PV auscultation sites at a sample rate of 4,000 Hz. Out of the 942 patients, 695 were classified as having no heart murmurs, 179 as having heart murmurs, and 68 as 'unknown'. Each signal was annotated by an expert based on whether it had a murmur or not; 'unknown' was assigned if the patient's status was unclear. The recordings for each patient ranged in duration from 5 to 65 seconds per channel. 
\subsubsection{Data Preparation and Labelling}
To ensure adequate heart sound information in every recording, we standardized the signal duration to 15 seconds. Longer signals were trimmed, and shorter ones were zero-padded. Additionally, we converted patient-wise labeling into sample-wise labeling as follows: if a patient’s label was "present", all samples from that patient were labeled as "present"; similarly, if the patient’s label was "absent", all their samples were labeled as "absent".
In this study, we did not use the data with unknown labels as we do not have any unknown labels in the main dataset from Bangladesh. Additionally, the presence and absence labels are used because the presence of a murmur may indicate CHD. Furthermore, we reserved 10\% of the patient data as "unseen" for testing to validate the transferability of the model. A detailed review of the dataset is provided in Table \ref{tab:PhysionetInfo}.
\begin{table*}[htbp]
\centering
\caption{Baseline Characteristics of Patients Included in the Study, Grouped by Valvular Murmur Status}
\label{tab:PhysionetInfo}
\begin{tabular}{|l|c|c|c|c|}
\hline
\textbf{Category} & \textbf{Overall} & \textbf{Absent} & \textbf{Present} & \textbf{Unknown} \\ 
                  & \textbf{n = 942} & \textbf{n = 695 (73.78\%)} & \textbf{n = 179 (19.00\%)} & \textbf{n = 68 (7.22\%)} \\ \hline
\textbf{Age} & & & & \\ 
Neonate & 6 (0.65\%) & 4 (0.58\%) & 1 (0.56\%) & 1 (1.47\%) \\ 
Infant & 126 (13.38\%) & 76 (10.94\%) & 25 (13.97\%) & 25 (36.76\%) \\ 
Child & 664 (70.49\%) & 495 (71.22\%) & 132 (73.74\%) & 37 (54.41\%) \\ 
Adolescent & 72 (7.64\%) & 53 (7.63\%) & 16 (8.94\%) & 3 (4.41\%) \\ 
Unlabeled & 74 (7.86\%) & 67 (9.64\%) & 5 (2.79\%) & 2 (2.94\%) \\ \hline
\textbf{Sex} & & & & \\ 
Female & 486 (51.59\%) & 355 (51.08\%) & 92 (51.40\%) & 39 (57.35\%) \\ \hline
\textbf{Height} & 108 (74-130) & 117 (93-134) & 111 (87-130) & 82 (66-123) \\ \hline
\textbf{Weight} & 18.55 (9.59-28.80) & 21.70 (13.60-32.50) & 18.55 (11.10-28.40) & 12.15 (7.70-25.60) \\ \hline
\textbf{BMI} & 16.32 (14.32-18.45) & 16.81 (15.20-19.10) & 16.10 (14.47-17.81) & 17.87 (15.88-20.00) \\ \hline
\textbf{Pregnant} & 70 (7.43\%) & 65 (9.35\%) & 3 (1.68\%) & 2 (2.94\%) \\ \hline
\end{tabular}
\end{table*}
\subsection{PhysioNet Challenge 2016 Dataset}
The PhysioNet Challenge 2016 dataset is another publicly available dataset containing more than 3,000 heart sound PCG signals. These dataset divided into five different subsets for validation. Experts annotated these signals as either normal or abnormal based on confirmed diagnoses. Furthermore, the signals were mostly collected from adult patients, with only one signal per patient. For our study, we used the validation set, which contains data from 300 patients, to test how our model generalizes to different populations and tasks.
\subsection{Preprocessing}
\subsubsection{Down-sampling}
The first step in the preprocessing stage was to downsample the PCG signals to conserve computational resources, as the signals were originally recorded at a sampling frequency of 4000 Hz. In this research, we downsampled all the signals to 800 Hz, after applying anti-aliasing filtering, which captures the majority of the heart sound frequencies.
\subsubsection{Normalization}
To maintain the signal within a consistent range, we applied z-score normalization to each audio file. This process adjusted the mean to zero and the standard deviation to one, ensuring consistency in the signal and facilitating further analysis.
\subsubsection{Base Model Architecture}
The base model used in this study is a CNN with a depth of 10 inception modules and residual connections. It starts with an input layer that accepts raw audio signals. Each inception module applies 1D convolutions with varying kernel sizes, along with max pooling, to capture multi-scale features from the input signals. The model includes residual connections after every few inception modules to improve gradient flow and learning efficiency. The architecture ends with a global average pooling layer to reduce the feature map dimensions to a fixed size, followed by a dense output layer with a sigmoid activation function for binary classification. This design leverages the combination of multi-scale feature extraction and residual learning to effectively process and classify PCG signals for detecting CHD.
\subsubsection{Model Initialization}
The base model was initially trained on a related task using the train subset of PhysioNet Challenge 2022 dataset to obtain pre-trained weights for model initialization. Instead of training the model from scratch, we utilized these pre-trained weights, allowing the model to continue learning and fine-tuning them during training. This approach significantly accelerated the convergence of the training process. The pre-trained weights, which encapsulate knowledge of murmurs in the signals, were integrated into the model, enhancing CHD detection accuracy by leveraging insights from the PhysioNet dataset to improve overall performance.
\subsubsection{Fine-tuning for CHD Detection}
After loading the pre-trained weights, the final layer of the base model is adjusted for CHD detection. This involves adding a new dense output layer with a softmax activation function to classify PCG signals into two target classes: CHD and Non-CHD. The rest of the model, which includes multiple inception modules with residual connections, is then fine-tuned. This fine-tuning process updates the previously learned feature representations to adapt to the new dataset, optimizing the model performance for the CHD detection task.
\subsubsection{Training Procedure}
The model is trained using categorical cross-entropy loss and uses an Adam optimizer with a learning rate schedule to enhance optimization. Additionally, class weights are computed to address the class imbalance, ensuring that the model does not overfit the dominant class (CHD) during training. To avoid overfitting, an early stopping callback is implemented based on the accuracy of the validation. If the validation performance does not improve over a predefined number of epochs, training is halted, and the model reverts to the best state. A learning rate scheduler is also used to gradually reduce the learning rate, allowing the model to fine-tune more effectively in later epochs.

\subsection{Model Training on PhysioNet Challenge 2022 Dataset}
In the first phase of this study, we trained a 1-dimensional CNN with the Inception Time architecture using the train subset of the PhysioNet Challenge 2022 dataset, which served as the source dataset for our approach. This training focused on detecting murmurs in audio signals and distinguishing between heart sounds with and without murmurs. The model learned features that are useful for identifying murmur-related heart sounds. Since the dataset contained three labels—'murmur present,' 'murmur absent,' and 'unknown'—we focused on two labels: 'murmur present' and 'murmur absent,' excluding the 'unknown' labels as they were not present in our main Bangladesh dataset. The challenge dataset was divided patient-wise into 90\% for training and 10\% for testing, allowing us to optimize the model while reserving a portion for performance evaluation.
\subsection{Model Training on main dataset}
In the second phase of this study, we trained the same model on the train subset of the main dataset from Bangladesh for CHD detection, utilizing the weights from the pre-trained model trained on the PhysioNet Challenge 2022 dataset. This approach allows the target model to benefit from the features learned in the source dataset, which involves detecting murmurs in heart sounds. Since murmurs often indicate heart abnormalities like CHD, the pre-trained model weights enable the target model to quickly recognize relevant patterns in the mian dataset, enhancing its performance in detecting CHD from heart sound signals. The block diagram of this study is shown in Figure \ref{fig1:block1}.
\subsection{Data imbalance}
The dataset used in this study has a CHD to non-CHD ratio of 63:37, highlighting a significant class imbalance. This imbalance was carefully maintained during the division of the dataset into training, validation, and testing sets to preserve the same ratio across each subset.

To address this issue in the context of heart sound samples from CHD and non-CHD patients, we incorporate the class weight function into the loss function. This approach ensures that the model prioritizes samples from underrepresented classes, as classes with higher weights contribute more to the loss function. The class weights were determined using the inverse frequency method, as defined in the equation below:
\[
W_i = \frac{N}{K \sum_{n=1}^{N} t_{ni}} \tag{1}
\]
Where $W_i$ is the class weight for class $i$, $N$ is the total number of samples, $K$ is the number of classes, and $t_{ni}$ indicates whether the $n^{th}$ sample belongs to the $i^{th}$ class.
\subsection{Evaluation metrics}
All evaluations were conducted on unseen test sets from the Bangladesh dataset and PhysioNet 2022 and also PhysioNet 2016, using patients data which were excluded from the model training.
Accuracy, Sensitivity, Specificity, and Area Under the Receiver Operating Characteristic Curve (AUROC) are used for each experiment to assess the efficacy of the proposed technique. The efficacy of detection algorithms is frequently evaluated using these metrics.
\section{Results}

\begin{figure*}[ht]
    \centering
    \begin{minipage}{0.5\textwidth}
        \centering
        \includegraphics[width=\linewidth]{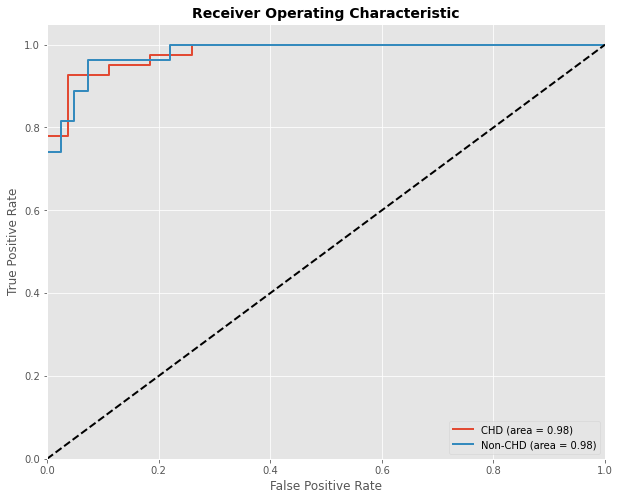}
    \end{minipage}%
    \begin{minipage}{0.5\textwidth}
        \centering
        \includegraphics[width=\linewidth]{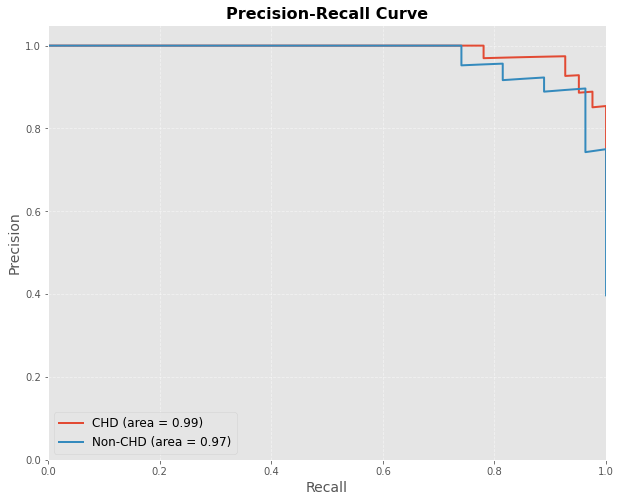}
    \end{minipage}
    \caption{Performance Evaluation on the Unseen Patient-Wise Split of the main Dataset: (a) Receiver Operating Characteristic (ROC) Curve (b) Precision-Recall Curve}
    \label{fig5:Results}
\end{figure*}
\begin{figure*}[ht]
    \centering
    \begin{minipage}{0.5\textwidth}
        \centering
        \includegraphics[width=\linewidth]{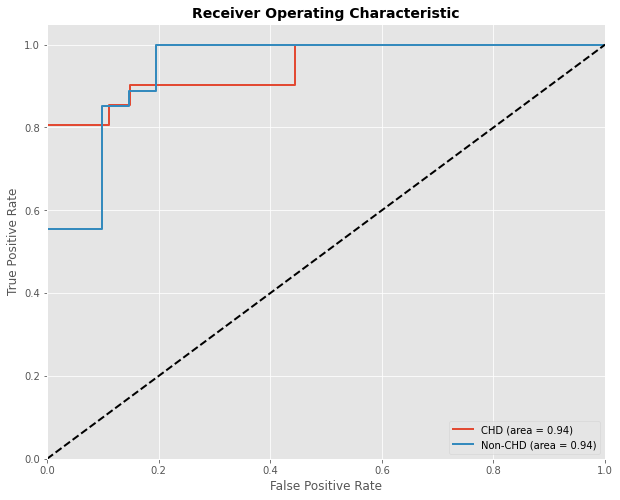}
    \end{minipage}%
    \begin{minipage}{0.5\textwidth}
        \centering
        \includegraphics[width=\linewidth]{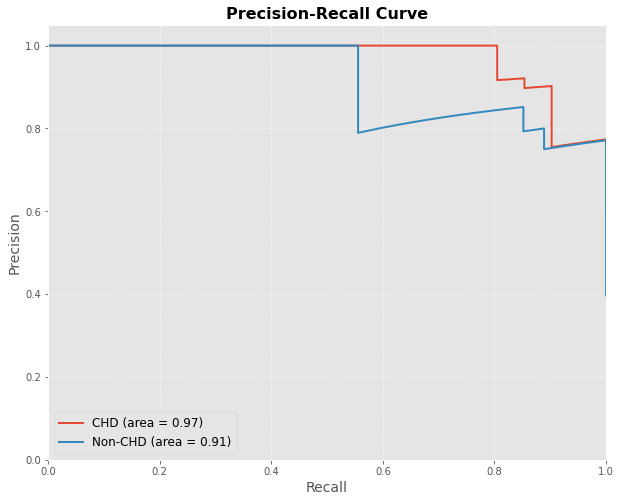}
    \end{minipage}
    \caption{Performance Evaluation on the Unseen Patient-Wise Split of the PhysioNet challenge 2022 dataset: (a) Receiver Operating Characteristic (ROC) Curve (b) Precision-Recall Curve}
    \label{fig4:Results}
\end{figure*}

\begin{table*}[h]
\centering
\footnotesize
\caption{Model Evaluation Across Different Datasets}
\begin{tabular}{|c|c|c|c|}
\hline
\multicolumn{4}{|c|}{\textbf{Testing on main Dataset from Bangladesh}} \\
\hline
\textbf{Accuracy (\%)} & \textbf{Sensitivity (\%)} & \textbf{Specificity (\%)} & \textbf{AUROC (\%)} \\
\hline
94.1 & 92.7 & 96.3 & 98.1 \\
\hline
\multicolumn{4}{|c|}{\textbf{Testing on Low-Quality Data from Bangladesh}} \\
\hline
\textbf{Accuracy (\%)} & \textbf{Sensitivity (\%)} & \textbf{Specificity (\%)} & \textbf{AUROC (\%)} \\
\hline
80.5 & 79.7 & 81.5 & 87.7 \\
\hline
\multicolumn{4}{|c|}{\textbf{Testing on PhysioNet Challenge 2022 Dataset}} \\
\hline
\textbf{Accuracy (\%)} & \textbf{Sensitivity (\%)} & \textbf{Specificity (\%)} & \textbf{AUROC (\%)} \\
\hline
91.6 & 92.2 & 88.9 & 94.4 \\
\hline
\multicolumn{4}{|c|}{\textbf{Testing on PhysioNet Challenge 2016 Dataset}} \\
\hline
\textbf{Accuracy (\%)} & \textbf{Sensitivity (\%)} & \textbf{Specificity (\%)} & \textbf{AUROC (\%)} \\
\hline
89.7 & 55.6 & 98.6 & 89.3 \\
\hline
\end{tabular}%
\label{tab:performance}
\end{table*}
\subsection{Evaluation on the main dataset}
The primary objective of this study was to detect CHDs in the Bangladesh dataset. Our model demonstrated balanced performance across all aspects, with an accuracy of 94.1\%, sensitivity of 92.7\%, specificity of 96.3\%, and AUROC of 98.1\%. Since we have PCG signals from four channels for each patient, the classification of CHD was determined by averaging the probabilistic outputs from the sound recordings taken from various channels. All results are presented in Table \ref{tab:performance}, and the ROC and Precision-Recall curves are shown in Figure \ref{fig5:Results}.
\subsection{Evaluation on the PhysioNet challenge 2022 dataset}
The train subset of PhysioNet Challenge 2022 dataset is used as the source dataset in our algorithm to leverage its pre-trained weights for CHD detection. We also tested the model on an unseen, separate patient split of this dataset, which demonstrated balanced performance in detecting murmurs in PCG signals, achieving an accuracy of 91.6\%, sensitivity of 92.2\%, specificity of 88.9\%, and an AUROC of 94.4\%. The results are presented in Table \ref{tab:performance}, and its ROC and Precision-Recall curves are shown in Figure \ref{fig4:Results}.

\begin{table*}[h]
\centering
\caption{Comparison Table}
\resizebox{\textwidth}{!}{%
\begin{tabular}{|l|c|c|c|c|c|}
\hline
\multicolumn{6}{|c|}{\textbf{PhysioNet Challenge 2022 (Top 2)}} \\
\hline
\textbf{Author (Year)} & \textbf{Dataset} & \textbf{Accuracy, \%} & \textbf{Sensitivity, \%} & \textbf{Specificity, \%} & \textbf{AUROC, \%} \\
\hline
{Yujia Xu \textit{et al.} (2022) \cite{xu2022hierarchical}} & {PhysioNet2022} & 83.8 & 94.4 & 65.5 & 92.2\\
\hline
{Hui Lu \textit{et al.} (2022) \cite{lu2022lightweight}} & {PhysioNet2022} & 88.2  & 95.5 & 75.5 & 86.6\\
\hline
\textbf{Author (Year)} & \textbf{Dataset} & \textbf{Accuracy, \%} & \textbf{Sensitivity, \%} & \textbf{Specificity, \%} & \textbf{AUROC, \%} \\
\hline
{Hassanuzzaman \textit{et al.} (2023) \cite{bib45}} & {Private} & 92 & \textbf{97} & 83 & 96 \\
\hline
\textbf{Our Work} & {Private} & \textbf{94} & \textbf{93} & \textbf{96} & \textbf{98} \\
\hline
\end{tabular}%
}
\label{tab:performance comparison}
\end{table*}
\subsection{Evaluation on the PhysioNet challenge 2016 dataset}
While PhysioNet Challenge 2016 data was not used for training, the proposed approach was tested on this completely unseen data, to assess the generalizability of our algorithm to various populations. While testing the model on this dataset, it underperformance in terms of sensitivity, achieving a sensitivity of 55.5\%, but demonstrated stronger performance regarding overall accuracy, specificity, and AUROC, with values of 89.7\%, 98.6\%, and 89.3\%, respectively. The low sensitivity may be attributed to the fact that the PhysioNet Challenge 2016 dataset predominantly includes adult patients,up to 60 years. In contrast, our model was trained on pediatric patient data, with ages from newborns to 21 years. Furthermore, the PhysioNet Challenge 2016 dataset, was not used as part of our training, and the labels may refer to other types of  heart diseases.
\subsection{Location-wise analysis}
\begin{figure*}[ht]
    \centering
    \includegraphics[width=0.7\textwidth]{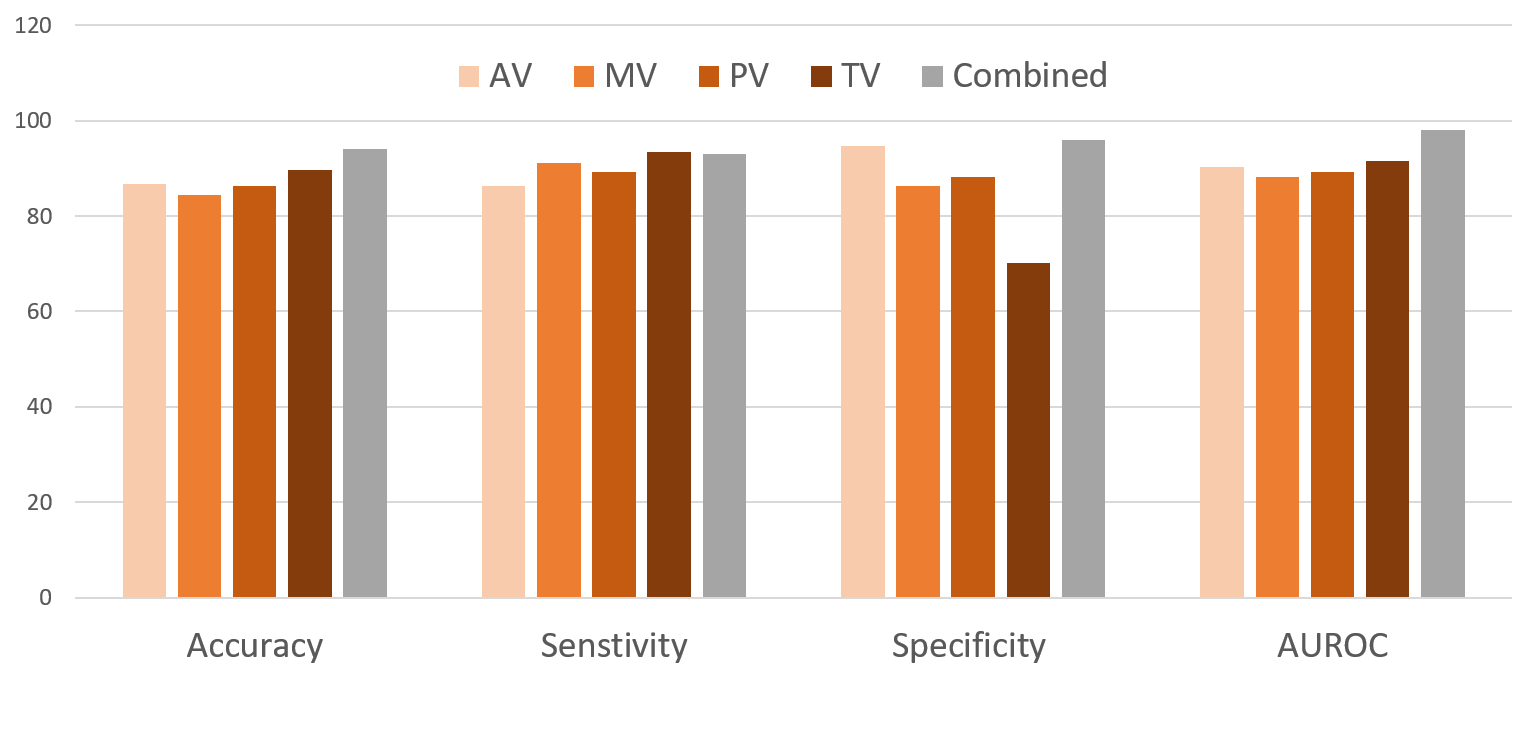}
    \caption{Model performance evaluation based on training with each individual PCG channel compared to the original combined PCG model. The channels include the aortic valve (AV), mitral valve (MV), pulmonary valve (PV), and tricuspid valve (TV). The performance of the combined model is represented by the final bar in each plot}
    \label{fig:location-wise}
\end{figure*}

To explore the significance of auscultation sites on chest, we trained the model on each site separately and compared its performance to the original combined PCG-based model. Notably, the highest accuracy of 85.68\% was achieved with the AV site, which also demonstrated a higher AUROC of 91.6\% compared to the other locations. Overall, the performance was maximized when using all channels combined as expected, however the performance of the model for single sites is still acceptable, as shown in Figure \ref{fig:location-wise}.
\subsection{Model performance on the low-quality data}
While testing our model on low-quality PCG signals, which were annotated by the cardiologists as unsatisfactory or very noisy, our model demonstrated balanced performance, with an accuracy of 80.5\%, sensitivity of 79.7\%, specificity of 81.5\%, and an AUROC of 87.7\%. The results are presented in Table \ref{tab:performance}.
\section{Comparison with previous study}
As research on CHD in LMICs is limited, we used the main Bangladesh dataset to evaluate the performance of the model. To compare our model with existing research, we ran the top two performing models from the PhysioNet Challenge 2022 on our dataset to provide a direct and more meaningful comparison. Additionally, we compared our model with other methods that utilized the same dataset. In comparison, our model demonstrates higher performance in terms of AUROC, specificity, and balanced accuracy, although it shows slightly lower sensitivity. Furthermore, the other methods exhibited poor performance in terms of specificity, which may place an extra burden on the healthcare system. All results are presented in Table \ref{tab:performance comparison}.
\section{Discussion}
Detecting CHD is particularly critical in infancy and childhood. The majority of detected murmurs for this age group indicate CHD \cite{Alkhodari2024}. Several studies have highlighted the importance of timely clinical assessment for CHD within the first few weeks of life, as early detection can significantly reduce morbidity and mortality in newborns. However, detecting CHD at an early stage can be challenging in LMICs due to the limited availability of echocardiography and trained personnel. Our study suggests that a digital stethoscope could be used as an alternative, with a dedicated deep learning software, to aid clinical decision support, as a powerful screening tool. The findings of this study have several important implications. Firstly, as shown in Table \ref{tab:performance comparison}, our study demonstrates a high performance on the main data collected in Bangladesh, which falls within the category of LMICs. This finding demonstrate the potential of our model to serve as a valuable screening tool in LMICs with large populations, where low-cost screening methods like this for CHD could be instrumental. Since our study focuses on a young patient cohort, achieving a CHD prediction accuracy of 94\% at an early stage through an automated method could significantly influence treatment plans, medication, and interventions \cite{chorba2021deep}. This, in turn, would enhance the quality of life for patients and minimize the risk of late or missed diagnoses.

The performance and efficacy of the proposed tool could open new avenues for implementing advanced techniques in clinical settings to support clinicians in decision-making. While fetal and neonatal echocardiography remains the gold standard for CHD diagnosis, the proposed tool offers a cost-effective and continuous screening alternative for resource-poor settings. Given that heart sounds are continuous acoustic waves, frequent monitoring through a simple yet powerful AI-driven tool could be highly beneficial, particularly in ongoing monitoring and assessments. While we tested our model on three databases, it would be interesting to assess its generalizability on additional databases before introducing it into clinical practice.

We evaluated our model using the PhysioNet Challenge 2022 and 2016 datasets to assess its transferability and generalizability to a new population, data collected from different devices, and for slightly different classification tasks. As shown in Table \ref{tab:performance}, the model demonstrated balanced performance when tested on the PhysioNet Challenge 2022 dataset, achieving an accuracy of 92\% and an AUROC value of 94\%. These results indicate that our model can serve as an effective screening tool in any LMIC without the need for additional training. Furthermore, the findings suggest that even a model trained on CHD classification in a specific population in Bangladesh is transferable for the different task of murmur detection in another population, even when using different devices. 

In the case of the PhysioNet Challenge 2016 dataset, which was only used as unseen data for testing, the model exhibited lower performance, which could be due to multiple factors. The dataset included a more diverse age group of both children and adults, it encompasses various heart diseases beyond CHDs, with more diverse heart valve disease and coronary artery disease. Additionally, the recordings in PhysioNet 2016 were collected using heterogeneous devices in both clinical and non-clinical settings, further contributing to the differences between the datasets. Another factor contributing to the model's poor performance on the PhysioNet Challenge 2016 dataset may be the difference in distributions, particularly the ratio of diseased to healthy populations. Nevertheless, the performance remains comparable to the top-performing models proposed for the challenge. Overall, our results indicate that the model is generalizable to new populations in LMICs.

To assess the location-specific significance of different auscultation sites on the chest for CHD detection, our analysis revealed that the TV channel is particularly important for identifying CHD, whereas the AV channel is more critical for the non-CHD group. As expected, the performance of our model is maximized when the data from all four locations are combined for the CHD detection task. However, even without precise placement and collection of data from the four locations, it is still possible to achieve over 85\% accuracy.

Since PCG signals are prone to noise, this poses critical concerns, particularly for infants and children, as factors like rubbing noises or crying can further compromise signal quality. In resource-limited settings, data is often collected by non-experts, which may introduce additional noise due to suboptimal placement of the stethoscope on the chest. When testing our model with unsatisfactory unseen data marked by experts as noisy or unsatisfactory, the model exhibited poorer performance compared to when tested with satisfactory data, as expected. While the model achieved an accuracy of 94\% on satisfactory quality data, it demonstrated only 80\% accuracy when evaluated on the unsatisfactory data. This highlights the expected decline in performance with lower quality inputs; however, it is important to note that the quality of the unsatisfactory quality data was such that cardiologists could not make any diagnoses based on these recordings, yet the algorithm was able to achieve an accuracy of 80\%, surpassing human expert perception.

Furthermore, we worked with PCG signals in their raw form, without human intervention or prior segmentation. This approach minimizes the dependence on expert annotations that might not always provide the necessary accuracy and annotation agreement. Overall, our analysis shows that the PCG signals, along with their extracted features, provide a rich dataset for building deep learning tools capable of processing large volumes of patient data, learning from hidden characteristics, and delivering automated medical predictions that can aid in clinical decision-making even for slightly different tasks. 

Despite the potential presented by our study, there are certain limitations to consider before applying our findings in a clinical setting. First, since we tested our model on two pediatric datasets and one mixed dataset including adult patients, it would be beneficial to validate our model on different age groups and additional external databases. Second, our model performed differently on unseen unsatisfactory data compared to satisfactory data, highlighting the potential need for a noise removal and signal quality assessment algorithms. Additionally, our algorithm may benefit from the segmentation of heart sounds, such as S1 and S2, as part of preprocessing. In this work, we limited the analysis to 15-second PCG signal segments, which may have affected performance, especially when signal quality fluctuates. To address this, incorporating the full length of recordings and integrating a signal quality assessment algorithm during preprocessing to select the highest quality segments of the recordings would enhance reliability. Moreover, we only performed binary classification of CHDs and did not further classify the CHD types. While this binary approach aids in screening and decision-making for patient referrals for further tests, it would be beneficial to explore multi-class classification in future work.

\section{Conclusion}
In this study, we developed and evaluated a deep learning model for the detection of CHD using PCG signals, emphasizing its applicability LMICs.  Our model demonstrated robust performance across various datasets, achieving an accuracy of 94\% on the primary Bangladesh dataset and maintaining a comparable performance on the PhysioNet Challenge public datasets. The findings highlight the potential of utilizing a digital stethoscope integrated with advanced AI algorithms as an effective screening tool for CHD, particularly in resource-limited settings where access to echocardiography is often limited. Our analysis revealed the significance of different auscultation sites in improving diagnostic accuracy and the model's resilience to varying data quality, showcasing its ability to outperform human expertise even in challenging conditions. Future work should focus on refining the model, exploring multi-class classification, and implementing additional noise mitigation strategies to maximize clinical utility.
\section*{References}
\bibliographystyle{IEEEtran}
\bibliography{IEEEabrv,references}

\end{document}